\documentclass[12pt]{iopart}

\usepackage{amsfonts}
\usepackage[english]{babel}
\usepackage{euscript}
\usepackage{bbm}
\usepackage{color}

\begin{document}

\title[Gibbons--Tsarev equation: symmetries, invariant solutions, and applications]%
{The Gibbons--Tsarev equation: symmetries, invariant solutions, and applications}

\author{Aleksandra Lelito, Oleg I. Morozov}

\address{Faculty of Applied
  Mathematics, AGH University of Science and Technology,
  \\
  Al. Mickiewicza 30,
  Cracow 30-059, Poland
\\
\vskip 7 pt
alelito{\symbol{64}}agh.edu.pl,
morozov{\symbol{64}}agh.edu.pl}


\begin{abstract}
In this paper we present the full classification of the
symmetry-invariant  solutions for the Gibbons--Tsarev equation.
Then we use these solutions to construct explicit expressions for reductions of Benney's moments
equations, to get solutions of Pavlov's equation, and to find integrable reductions of the
Ferapontov--Huard--Zhang system, which describes implicit two-phase solutions of the dKP equation.
\end{abstract}

\maketitle
\section{Introduction}

The Gibbons-Tsarev equation
\begin{equation}\label{Gibbons_Tsarev_eq}
u_{yy}=(u_y+y)\,u_{xx}-u_x\,u_{xy}-2
\end{equation}
has been widely known since it arose in \cite{GibbonsTsarev1996} as a special case of a reduction of Benney's
moment equations, \cite{Benney1973},
\begin{equation}
A_{n,t}+A_{n+1,x}+n\,A_{n-1}\,A_{0,x} = 0,
\qquad n \in \mathbb{N} \cup \{0\}.
\label{Benney_chain}
\end{equation}
Namely, suppose that $A_2$ and $ A_3$ depend functionally on $p \colon = A_0$ and $q\colon = A_1$, that is,
$A_2=R(p,q)$, $A_3=S(p,q)$ for some functions $P$ and $S$. Then for all $n \ge 4$ moments $A_n$  also depend
functionally on $A_0$ and $A_1$, $A_n=Q_n(p,q)$, where all the functions $Q_n$ may be expressed recurrently in
terms of $R$ and $S$. Substituting for $A_2=R(p,q)$, $A_3=S(p,q)$ into (\ref{Benney_chain}) yields an
over-determined system
\begin{equation}
\left\{
\begin{array}{lcl}
S_q&=&R_p+R_q^2,
\\
S_p&=&R_q\,(R_p+p)-2\,q,
\end{array}
\right.
\label{reduced_Benney_eqns}
\end{equation}
which is compatible whenever $R_{pp}=(R_p+p)\,R_{qq}-R_q\,R_{pq}-2$. This equation coincides with (\ref{Gibbons_Tsarev_eq}) after renaming $(q,p,R) \mapsto (x,y,u)$.

This origin of the Gibbons--Tsarev equation connects it directly with a model (also pre\-sen\-ted in
the above-mentioned Benney's work), which is
meant to describe behavior of long waves on
a shallow, inviscid and incompressible fluid. The Gibbons--Tsarev equation
also arises in integrable models on algebraic curves, \cite{MarikhinSokolov}.
In \cite{Kaptsov2002,KaptsovSchmidt} the method of differential constraints was applied to find solutions of the Gib\-bons--Tsarev
equation that are expressible in terms of solutions of Painlev{\'e} equ\-a\-ti\-ons.

In this paper we use the methods of group analysis of differential equations, see, e.g., \cite{Olver}, to find solutions of the Gibbons--Tsarev
equation that are invariant with respect to its symmetries. The research, as usually when Lie theory is applied, is performed in an algorithmic
way, which involves reduction of the primary equation into equation in less independent variables than the primary one. The basic method
includes arbitrary choice of symmetries, but one may confine this choice in some sense with the help of the adjoint representation of the symmetry algebra. In order to examine the problem of finding group-invariant solutions in as much systematic way as possible, we applied the method
based on searching for an optimal system of one-dimensional subalgebras of the symmetry algebra of equation (\ref{Gibbons_Tsarev_eq}).  Hence, every other group-invariant solution can be derived from one of the solutions we obtained.

As an immediate application of the invariant solutions of equation (\ref{Gibbons_Tsarev_eq}) we get explicit forms for four
reductions of Benney's moments equations. Two further applications are the following. First, as it was shown in
\cite{BKMV2014}, equation (\ref{Gibbons_Tsarev_eq}) arises as a symmetry reduction of equation \cite{Pavlov2003,Dunajski2004}
\begin{equation}
u_{yy} = u_{tx}+u_y\,u_{xx}-u_x\,u_{xy}.
\label{Pavlov_eq}
\end{equation}
Thus, solutions to equation (\ref{Gibbons_Tsarev_eq}) provide solutions to equation (\ref{Pavlov_eq}).
Second, the change of variables
\begin{equation}
z=u+\case{1}{2}\,y^2
\label{v_u_substitution}
\end{equation}
transforms (\ref{Gibbons_Tsarev_eq}) to the first equation of the system
\begin{equation}
\left\{
\begin{array}{lcl}
z_{yy} &=& z_y\,z_{xx} - z_x\,z_{xy} - 1,
\\
w_{yy} &=& z_y\,w_{xx} - z_x\,w_{xy}.
\end{array}
\right.
\label{FHZ_system}
\end{equation}
This system was shown in \cite{FHZ2012} to produce two-phase solutions for the
dispersionless Kadomtsev--Petviashvili equation (dKP). Namely,  if functions $P(r,s)$, $Q(r,s)$ satisfy
\begin{equation}
\left\{
\begin{array}{lcl}
P_{ss} &=& P_s\,P_{rr} - P_r\,P_{rs} - 1,
\\
Q_{ss} &=& P_s\,Q_{rr} - P_r\,Q_{rs},
\end{array}
\right.
\label{FHZ_PQ_system}
\end{equation}
then the system
\begin{equation}
\left\{
\begin{array}{lcl}
Q_{r} &=& x+ t\,(r+P_r),
\\
Q_{s} &=& y + t\,P_s,
\end{array}
\right.
\label{PQ_system}
\end{equation}
implicitly defines a solution $r(t,x,y)$, $s(t,x,y)$ to the system
\[
\left\{
\begin{array}{lcl}
r_t &=& r\,r_x+ s_y,
\\
r_y &=& s_x,
\end{array}
\right.
\]
which is equivalent to the dKP equation
\begin{equation}
r_{yy} = r_{tx} - (r\,r_x)_x.
\label{pdKP}
\end{equation}
Each solution to (\ref{Gibbons_Tsarev_eq}) yields by substituting (\ref{v_u_substitution}) into (\ref{FHZ_system}) a
linear equation for $w$. We analyse symmetries of the obtained linear equations. Their corresponding
reductions appear to be ordinary differential equations equivalent to Airy's equation,
\[
v_{xx} = x\,v,
\]
Weber's equation
\[
v_{xx} = \left(\frac{1}{4}\,x^2+\lambda\right)\,v,
\]
Whittaker's equation
\[
v_{xx} = \left(\frac{1}{4}-\frac{\kappa}{x}+\frac{4\,\mu^2-1}{4\,x^2}\right)\,v,
\]
and Bessel's equation
\[
v_{xx} = \left(\frac{1}{4}+\frac{4\,\mu^2-1}{4\,x^2}\right)\,v,
\]
see, e.g., \cite{WhittakerWatson,AbramowitzStegun}.  While Airy's equation is not integrable in quadratures, \cite{Kaplansky},
for Weber's equation, Whittaker's equation and Bessel's equation there exist an infinite number of values of the
parameters $\lambda$, $\kappa$, $\mu$ such that those equations are integrable, see \cite{Ritt1948},
\cite{Kovacic1986} and \cite{RamisMartinet}. Hence, we obtain an infinite number of cases when system
(\ref{FHZ_system}) is integrable in quadratures. While the corresponding solutions to (\ref{FHZ_PQ_system}),
(\ref{PQ_system}) describe two-phase solutions for Eq. (\ref{pdKP}), their final form appears to be too
complicated to write it explicitly.

\section{The symmetry algebra of the Gibbons--Tsarev equation}

Roughly speaking, a \textit{symmetry group} of an equation
\begin{equation}\label{eq:F}
F(x,y,u,u_x,u_y,u_{xy},u_{xx},u_{yy})=0,
\end{equation}
is a local group $G$ of transformations $g$ acting on some open subset of the space of independent and dependent
variables $X\times Y\times U$, which transform solutions of the equation into solutions (for the strict definition
see \cite{Olver}). There is a one-to-one cor\-res\-pon\-den\-ce between symmetry group and its
\textit{infinitesimal generator}, which is a vector field of the form:
\[
V=\xi_1(x,y,u)\frac{\partial}{\partial x}+\xi_2(x,y,u)\frac{\partial}{\partial y}
+\eta(x,y,u)\frac{\partial}{\partial u}.
\]
Every infinitesimal generator has its \textit{characteristic function}, defined as $Q=\eta-\xi_1\,u_x-\xi_2\,u_y$,
which is very useful from the computational point of view. In this paper, by symmetry we mean either a
characteristic or a corresponding vector field, depending on a context. Finally, a \textit{symmetry algebra} is a
set of infinitesimal generators of symmetries, closed with respect to commutator $[\cdot,\cdot]$.
For any two vector fields $V_1$, $V_2$, their commutator is defined as $[V_1,V_2]:=V_1\circ V_2-V_2 \circ V_1$.

If $u=f(x,y)$ is a solution of (\ref{eq:F}), then it is $G$-invariant if for any group trans\-for\-ma\-ti\-on $g$ the
transformed function $(g\cdot f)(x,y)$ is a solution to Eq. (\ref{eq:F}), too.
In what follows the notation will be  widely based on the one adopted in \cite{Olver}.

\subsection{Symmetry algebra}

With the help of Jets software, \cite{Jets}, we found the symmetry algebra of equation (\ref{Gibbons_Tsarev_eq}), which is
presented in the following table.
\[
\begin{array}{c|l|l}\label{tab:sym}
\mathrm{Symmetry} & \mathrm{Characteristic} & \mathrm{Vector\quad field} \\ \hline
\phi_1 & -y\,u_x+2\,x & y\,\frac{\partial}{\partial x}+2\,x\,\frac{\partial}{\partial u} \\
\phi_2 & -x\,u_x-\frac{2}{3}\,y\,u_y+\frac{4}{3}\,u & x\,\frac{\partial}{\partial x}+\frac{2}{3}\,y\,\frac{\partial}{\partial y}+\frac{4}{3}\,u\,\frac{\partial}{\partial u} \\
\phi_3 & -u_x & \frac{\partial}{\partial x} \\
\phi_4 & -u_y-y & \frac{\partial}{\partial y}-y\,\frac{\partial}{\partial u} \\
\phi_5 & 1 & \frac{\partial}{\partial u}
\end{array}
\]
The commutator table of this Lie algebra is the following:
\[
\begin{array}{c|c|c|c|c|c}
\rule[-1ex]{0pt}{2.5ex}  & \phi_1 & \phi_2 & \phi_3 & \phi_4 & \phi_5 \\
\hline
\rule[-1ex]{0pt}{2.5ex} \phi_1 & 0 & \frac{1}{3}\phi_1 & -2\phi_2 & -\phi_3 & 0 \\
\hline
\rule[-1ex]{0pt}{2.5ex} \phi_2 & -\frac{1}{3}\phi_1 & 0 & -\phi_3 & -\frac{2}{3}\phi_4 & -\frac{4}{3}\phi_5 \\
\hline
\rule[-1ex]{0pt}{2.5ex} \phi_3 & 2\phi_2 & \phi_3 & 0 & 0 & 0 \\
\hline
\rule[-1ex]{0pt}{2.5ex} \phi_4 & \phi_3 & \frac{2}{3}\phi_4 & 0 & 0 & 0 \\
\hline
\rule[-1ex]{0pt}{2.5ex} \phi_5 & 0 & \frac{4}{3}\phi_5 & 0 & 0 & 0 \\
\end{array}
\]
Note that $\phi_2$ is a scaling symmetry, while $\phi_3$ and $\phi_5$ denote invariance of the set of solutions
with respect to translations of $x$ and $u$.

\subsection{Adjoint representation}

The full symetry group of Gibbons--Tsarev equation is generated by five one-dimensional subgroups whose generators are presented in table
(\ref{tab:sym}). Reduction with respect to one of these 1-dimensional subgroups gives us an equation in $2-1=1$
variables. Any linear combination of symmetries is again a symmetry and it brings new reduction, which makes the
task of finding all group-invariant solutions very tedious. However, it is easy to check, that if $f(x,y)$ is a
$G$-invariant solution, then $(h\cdot f)(x,y)$ is $h\,G\,h^{-1}$-invariant. This observation indicates the need
of finding a set of solutions, which are invariant only to subgroups laying on separate orbits with respect to
conjugation. As usual, we will operate on vector fields rather than subgroups of transformations themselves.
The {\it adjoint representation} is defined as follows.
For a given vector $V$ from a Lie algebra  denote by $Ad_{\epsilon\, V}$  a linear map on the Lie
algebra, which is defined for every vector $W$ from the Lie algebra as follows:
\[
Ad_{\epsilon\, V}\,W:=W-\epsilon\,[V,\,W]
+\frac{\epsilon^2}{2!}\,[V,\,[V,\,W]]-\frac{\epsilon^3}{3!}\,[V,\,[V,\,[V,\,W]]]+\cdots
\]
The adjoint representation has a useful property of transforming vector $W$
generating a subgroup $G_W$ to the vector $Ad_{\epsilon\, V}\,W$ generating subgroup $h\,G_W\,h^{-1}$, where
$h=\exp(\epsilon\, V)$. The adjoint representation for the symmetry algebra of the  Gibbons--Tsarev equation
is presented in the following table. The $(i,j)$-th entry is $Ad_{\epsilon\, \phi_i}\,\phi_j$.
\[
\fl
\begin{array}{c|ccccc}
Ad_{\epsilon\, \phi_i}\,\phi_j & \phi_1 & \phi_2 & \phi_3 & \phi_4 & \phi_5 \\
\hline
\phi_1 & \phi_1 & \phi_2+\frac{\epsilon}{3}\,\phi_1 & \phi_3-2\,\epsilon\,\phi_5 & \phi_4-\epsilon\,\phi_3+\epsilon^2\,\phi_5 & \phi_5 \\
\phi_2 & e^{-\frac{1}{2}\,\epsilon}\,\phi_1 & \phi_2 & e^{-\epsilon}\,\phi_3 & e^{-\frac{2}{3}\,\epsilon}\,\phi_4 & e^{-\frac{4}{3}\,\epsilon}\,\phi_5 \\
\phi_3 & \phi_1+2\,\epsilon\,\phi_5 & \phi_2+\epsilon\,\phi_3 & \phi_3 & \phi_4 & \phi_5 \\
\phi_4 & \phi_1+\epsilon\,\phi_3 & \phi_2+\frac{2}{3}\,\epsilon\,\phi_4 & \phi_3 & \phi_4 & \phi_5 \\
\phi_5 & \phi_1 & \phi_2+\frac{4}{3}\,\epsilon\,\phi_5 & \phi_3 & \phi_4 & \phi_5
\end{array}
\]
The following lemma presents an optimal system of one-dimensional subalgebras for  the symmetry algebra of the Gibbons--Tsarev equation, by which a list of vectors generating conjugacy inequivalent one-parameter subgroups is meant, \cite[\S~3.3]{Olver}.

\vskip 5 pt
\noindent
{\bf Lemma.}
{\it The optimal system of one-dimensional subalgebras consists of the subalgebras spanned by the following vectors:}
$\phi_1$, $\phi_2$, $\phi_3$, $\phi_4+\alpha\,\phi_1$, $\phi_4+\alpha\,\phi_5$, where $\alpha$ is an arbitrary
constant.

\vskip 5 pt
\noindent
Proof is  obtained by a standard computation, see, e.g., \cite[\S~3.3]{Olver}.

\section{Reductions and invariant solutions}

In this section we find solutions of equation (\ref{Gibbons_Tsarev_eq}) that are invariant with respect to the optimal system obtained in the above lemma. We use the method described, e.g., in \cite[\S~3.1]{Olver}.

\subsection{Reduction with respect to $\phi_1$}

The $\phi_1$-invariant solutions of the Gibbons--Tsarev equation satisfy (\ref{Gibbons_Tsarev_eq}) and
\[
\phi_1=-y\,u_x+2\,x=0.
\]
Solving the last equation for $u_x$ and integrating gives $u=x^2\,y^{-1} + W(y)$.
Substituting this to (\ref{Gibbons_Tsarev_eq}) and solving for unknown function $W(y)$ yields
\begin{equation}
u=\frac{x^2}{y}+\beta\,y^3+\gamma,
\label{phi_1_invariant_solution}
\end{equation}
where $\beta$, $\gamma$ are arbitrary constants.

\subsection{Reduction with respect to $\phi_2$}

The $\phi_2$-invariant solutions of the Gibbons--Tsarev equation satisfy (\ref{Gibbons_Tsarev_eq}) and
\[
\phi_2=-x\,u_x-\frac{2}{3}\,y\,u_y+\frac{4}{3}\,u=0.
\]
Solving this we get $u=x^{4/3}\,v(\zeta)$ with $\zeta=y\,x^{-2/3}$. Inserting the outcome into (\ref{Gibbons_Tsarev_eq}) yields the ordinary differential equation
\begin{equation}
v_{\zeta\zeta} = \frac{2\,(\zeta\,v_\zeta^2+(2\,v+3\,\zeta^2)\,v_\zeta-2\,\zeta\,v+9)}{8\,\zeta\,v+4\,\zeta^3-9}.
\label{ode_phi_2}
\end{equation}
The point symmetries of this equation are trivial, so the methods of group analysis can not be applied to its
integration. The general solution to (\ref{ode_phi_2}) may be extracted from results of
\cite{PavlovTsarev2014}. It has the parametric form
\begin{equation}
\left\{
\begin{array}{lcl}
v  &=&\displaystyle{-\frac{3^{2/3}}{2\,(1+\epsilon_1+\epsilon_2)^{1/3}} \cdot
\frac{P_2(t)}{(P_3(t))^{2/3}}},
\\
\zeta &=&\displaystyle{-\frac{3^{4/3}}{2\,(1+\epsilon_1+\epsilon_2)^{2/3}} \cdot
\frac{P_4(t)}{(P_3(t))^{4/3}}}
\end{array}
\right.
\label{solution_to_ode_phi_2}
\end{equation}
with
\begin{eqnarray}
\fl
P_2(t)&=&(\epsilon_1+\epsilon_2\,t)^2+\epsilon_1+\epsilon_2\,t^2,
\nonumber \\
\fl
P_3(t)&=&(\epsilon_1+\epsilon_2\,t)^3-\epsilon_1-\epsilon_2\,t^3,
\nonumber \\
\fl
P_4(t)&=&(1+2\,(\epsilon_1+\epsilon_2))\,(\epsilon_1+\epsilon_2\,t)^4
+\epsilon_2\,(2\,(1+\epsilon_1-\epsilon_2^2)+\epsilon_2)\,t^4
-4\,\epsilon_1\,\epsilon_2^2\,t^3
\nonumber \\
\fl
&&-2\,\epsilon_1\,\epsilon_2\,(1+\epsilon_1+\epsilon_2)\,t^2
-4\,\epsilon_1^2\,\epsilon_2\,t
+\epsilon_1\,(\epsilon_1+2\,(1+\epsilon_2-\epsilon_1^2)),
\label{polynoms_solution_to_ode_phi_2}
\end{eqnarray}
where $\epsilon_1$ and $\epsilon_2$ are arbitrary constants and $t$ is a parameter.
Since
\[
\fl
\mathrm{det}\left(
\begin{array}{ll}
\displaystyle{\frac{\partial v}{\partial \epsilon_1}} & \displaystyle{\frac{\partial v}{\partial \epsilon_2}}
\\
\displaystyle{\frac{\partial v_\zeta}{\partial \epsilon_1}}& \displaystyle{\frac{\partial v_\zeta}{\partial \epsilon_2}}
\end{array}
\right)
=
\frac{3^{5/3}(1+\epsilon_1+\epsilon_2)^{2/3}(t-1)^2(\epsilon_1+(1+\epsilon_2)t)^2(1+\epsilon_1+\epsilon_2 t)^2}
{8\,(P_3(t))^{8/3}}
\not \equiv 0,
\]
system (\ref{solution_to_ode_phi_2}), (\ref{polynoms_solution_to_ode_phi_2}) indeed defines the general solution to
Eq. (\ref{ode_phi_2}). This fact was not proved in \cite{PavlovTsarev2014}.  This solution is very complicated, so
we will not use it in the constructions of Section \ref{Applications_Section}.

\subsection{Reduction with respect to $\phi_3$}

For  $\phi_3$-invariant solutions of the Gibbons--Tsarev equation we have
\[
\phi_3=-u_x=0,
\]
so they do not depend on $x$ and thus satisfy $u_{yy}=-2$. Hence these solutions are of the form
\begin{equation}
u=-y^2+\beta\,y+\gamma
\label{phi_3_invariant_solution}
\end{equation}
with $\beta, \gamma =\mathrm{const}$.

\subsection{Reduction with respect to $\phi_4+\alpha\phi_1$}

Solutions of the Gibbons--Tsarev equation that are invariant w.r.t. $\phi_4+\alpha\phi_1$ satisfy (\ref{Gibbons_Tsarev_eq}) and
\[
\phi_4+\alpha\,\phi_1=-\alpha\, y\,u_x-u_y-y+2\,\alpha\, x=0.
\]

When $\alpha\neq 0$, we solve this equation for $u_x$, substitute the output
into (\ref{Gibbons_Tsarev_eq})
and obtain the reduced equation
\[
u_{yy} =\frac{2\,(x-\alpha \,y^2)}{y\,(2\,x-\alpha\, y^2)}\,u_y-\frac{4\,\alpha\, x^2}{y\,(2\,x-\alpha\, y^2)}.
\]
This is a linear ordinary differential equation with
$x$ treated as a parameter. So\-lu\-ti\-ons of this equation are of the form
\[
u=2\,\alpha \,x\,y-\frac{2}{3}\,\alpha^2\,y^3
+W_1(x)\cdot \vert 2\,x - \alpha\, y^2 \vert^{\frac{3}{2}}
+W_2(x),
\]
where $W_1(x)$ and $W_2(x)$ are arbitrary (smooth) functions of $x$.
By substituting this solution to (\ref{Gibbons_Tsarev_eq}) we obtain that $W_1(x)=\beta=\mathrm{const}$
and $W_2(x)=-\alpha^{-1}\,x+\gamma$, $\gamma=\mathrm{const}$. Finally, solution invariant with respect to
symmetry $\phi_4+\alpha\,\phi_1$ is of the form:
\begin{equation}
u=(2\,\alpha \,y-\alpha^{-1})\,x-\frac{2}{3}\,\alpha^2\,y^3
+\beta\,\vert\alpha\, y^2-2\,x\vert^{\frac{3}{2}}
+\gamma.
\label{phi_4_plus_alpha_phi_1_invariant_solution}
\end{equation}

When $\alpha = 0$, we have $u_y=-y$, so $u= -\case{1}{2}\,y^2+ W_1(x)$. But substituting for this into (\ref{Gibbons_Tsarev_eq}) gives a contradiction.

\subsection{Reduction with respect to $\phi_4+\alpha \,\phi_5$}

Solutions of the Gibbons--Tsarev equation that are invariant w.r.t. $\phi_5+\beta\phi_4$ satisfy (\ref{Gibbons_Tsarev_eq}) and
\[
\phi_4+\alpha\,\phi_5=-u_y- y+\alpha=0.
\]
This gives
$u=-\case{1}{2}\,y^2+\alpha\,y+W(x)$. Substituting to (\ref{Gibbons_Tsarev_eq}) and solving for $W(x)$
gives the solution of the form
\begin{equation}
u=\frac{1}{2\,\alpha}\,x^2-\frac{1}{2}\,y^2+\alpha\,y+\beta\,x+\gamma
\label{phi_4_plus_alpha_phi_5_invariant_solution}
\end{equation}
with $\beta$, $\gamma = \mathrm{const}$.

\section{Applications}
\label{Applications_Section}

\subsection{Reductions of Benney's moments equation}

Renaming  $(q,p,R) \mapsto (x,y,u)$ in system  (\ref{reduced_Benney_eqns}) and substituting for a solution of equation (\ref{Gibbons_Tsarev_eq})
in\-to the resulting system
\[
\left\{
\begin{array}{lcl}
S_x&=&u_y+u_x^2,
\\
S_y&=&u_x\,(u_y+y)-2\,x,
\end{array}
\right.
\]
we obtain a compatible system for $S$. This system has the following solutions that cor\-res\-pond to the invariant solutions
(\ref{phi_1_invariant_solution}),
(\ref{phi_3_invariant_solution}),
(\ref{phi_4_plus_alpha_phi_1_invariant_solution}),
(\ref{phi_4_plus_alpha_phi_5_invariant_solution})
of the Gibbons--Tsarev equation, res\-pec\-ti\-ve\-ly:
\[
S ={\frac {{x}^{3}}{{y}^{2}}}+3\,\beta\,x\,y^{2}+\delta,
\]
\[
S=(\beta-2\,y)\,x+\delta,
\]
\[
S = \frac{\beta\,(3\,\alpha^2\,y-2)}{\alpha}\,\vert\alpha\, y^2-2\,x\vert^{\frac{3}{2}}
-\case{1}{4}\,\alpha^2\,(4\,\alpha+9\,\beta^2)\,y^4
-4\,x\,y+\frac{1}{\alpha^2}\,x
\]
\[
\qquad\qquad
+\case{4}{3}\,\alpha\,y^3+(\alpha-9\,\beta^2)\,x^2
+\frac{1}{2\,\alpha}\,(18\,\alpha^2\,\beta^2\,x+4\,\alpha^3\,x-1)\,y^2
+\delta,
\]
\[
S =\frac{x^3}{3\,\alpha^2}+\frac{\beta\,x^2}{\alpha}+(\alpha+\beta^2-y)\,x+\alpha\,\beta\,y+\delta,
\]
where $\delta$ is an arbitrary constant.

\subsection{Solutions to  equation (\ref{Pavlov_eq})}

Equation (\ref{Pavlov_eq})  has solutions of the form
\begin{equation}\label{eq:Pvlvsol}
u(t,x,y)=v(\tau,y)-2\,t\,x-t^2\,y,
\end{equation}
where $\tau=x+t\,y$ and function $v(\tau,y)$ is a solution of equation (\ref{Gibbons_Tsarev_eq}) with $x$ replaced by $\tau$.
Since we know four explicit solutions
(\ref{phi_1_invariant_solution}),
(\ref{phi_3_invariant_solution}),
(\ref{phi_4_plus_alpha_phi_1_invariant_solution}),
(\ref{phi_4_plus_alpha_phi_5_invariant_solution})
of the Gibbons--Tsarev equation, after substituting them into
(\ref{eq:Pvlvsol}) we obtain four explicit solutions of equation (\ref{Pavlov_eq}). They are, respectively,
\[
u=\frac{x^2}{y}+\beta\,y^3+\gamma,
\]
\[
u=-y^2+\beta\,y-2\,t\,x-t^2\,y+\gamma,
\]
\[
u=(2\,\alpha \,y-2\,t-\alpha^{-1})\,(x
+t\,y)
-\frac{2}{3}\,\alpha^2\,y^3
+\beta\,\vert  2\,x   - \alpha\, y^2 +t\,y\vert^{\frac{3}{2}}
+t^2\,y+\gamma,
\]
\[
u=\frac{1}{2\,\alpha}\,(x+t\,y)^2
-\frac{1}{2}\,y^2+\alpha\,y+\beta\,(x+t\,y)-2\,t\,x-t^2\,y+\gamma.
\]


\subsection{Reductions of the Ferapontov--Huard--Zhang system}

In this section we study solutions of system (\ref{FHZ_system}) that correspond to the obtained solutions
(\ref{phi_1_invariant_solution}),
(\ref{phi_3_invariant_solution}),
(\ref{phi_4_plus_alpha_phi_1_invariant_solution}),
(\ref{phi_4_plus_alpha_phi_5_invariant_solution})
of the Gibbons--Tsarev equation.
Each solution of (\ref{Gibbons_Tsarev_eq}) yields by sub\-sti\-tu\-ting (\ref{v_u_substitution}) into
(\ref{FHZ_system}) a linear equation for $w$. Any linear equation admits trivial sym\-met\-ri\-es, that is,
symmetries of the form $w_0 \,\frac{\partial}{\partial w}$, where $w_0$ is a (fixed) arbitrary solution of the
equation. We consider nontrivial symmetries of the obtained linear equations. The\-se symmetries allow one to
reduce their equations to ordinary differential equations. For each one of these ODEs we indicate all the cases
when the ODE is integrable in quadratures.

\subsubsection{Solution (\ref{phi_1_invariant_solution})}
\hspace {1 pt}
\vskip 2 pt
\noindent
For solution (\ref{phi_1_invariant_solution}) the second equation of system (\ref{FHZ_system}) takes the form
\[
w_{yy} = (3\,\beta\,y^2+y-x^2\,y^{-2})\,w_{xx}-2\,x\,y^{-1}\,w_{xy}.
\]
After the change of variables  $x=\tilde{x}\,\tilde{y}$, $y=\tilde{y}$,
$w=\tilde{w}$  and  dropping tildes the last equation acquires the form $w_{yy} = (3\,\beta\,y+1)\,y^{-1}\,w_{xx}$. This equation has a nontrivial symmetry $w_x -\lambda\,w$, where $\lambda$ is an arbitrary constant. The corresponding reduction $w=e^{\lambda\,x}\,v(y)$ gives
an ODE
$v_{yy} = \lambda^2\,(3\,\beta\,y+1)\,y^{-1}\,v$.
After the scaling  $\tilde{y} = 2\,\sqrt{3}\,\lambda\,\beta^{1/2}\,y$ and dropping tildes we have  Whittaker's
\begin{equation}
v_{yy} = \left(\frac{1}{4}-\frac{\kappa}{y} \right)\,v
\label{ode_1}
\end{equation}
with $\kappa = -\case{1}{6}\,\sqrt{3}\,\lambda\,\beta^{1/2}$.
From results of \cite{RamisMartinet} it follows that equation (\ref{ode_1}) is integrable in quadratures whenever  $\kappa \in \mathbb{Z}$.
Therefore for each choice of $\beta$ there exists an infinite number of values for $\lambda$ such that equation (\ref{ode_1}) is integrable in quadratures.

\subsubsection{Solution (\ref{phi_3_invariant_solution})}
\hspace {1 pt}
\vskip 2 pt
\noindent
Without loss of generality  it is possible to put $\beta = 0$ in solution (\ref{phi_3_invariant_solution}).
Then we have
\begin{equation}
w_{yy} = -y\,w_{xx}.
\label{pde_4_3_2}
\end{equation}
The nontrivial symmetries of this equation are the following: $\psi_1 =w_x-\lambda\,w$ with $\lambda = \mathrm{const}$,
$\psi_2 = 3\,x\,w_x+2\,y\,w_y$, and $\psi_3 = 12\,x\,y\,u_y+3\,x\,w-(4\,y^3-9\,x^2)\,u_x$.

The $\psi_1$-invariant solution of equation (\ref{pde_4_3_2}) is of the form  $w=e^{\lambda\,x}\,v(y)$, where $v$ satisfies
$v_{yy} = -\lambda^2\,y\,v$. After rescaling  $y=-\lambda^{2/3}\,\tilde{y}$ and  dropping tildes the last equation acquires the form of
Airy's equation
\[
v_{yy} = y\,v,
\]
which is not  integrable in quadratures, \cite{Kaplansky}.

The $\psi_2$-invariant solution of equation (\ref{pde_4_3_2}) is of the form  $w=v(\eta)$ with $\eta=x\,y^{-3/2}$, where  $v$ is a solution of
equation
\[
v_{\eta\eta} = -\frac{10\,\eta^2}{4\,\eta^3+9}\,v_\eta.
\]
The general solution of this equation is
\[
U = c_1+c_2\,\int\,\frac{d\eta}{(4\,\eta^3+9)^{5/6}}.
\]
The last integral can not be expressed in elementary functions, \cite{Ritt1948}.

For a $\psi_3$-invariant solution we have
\[
w=\frac{y}{(4\,y^3+9\,x^2)^{5/6}}\,v(\sigma),
\qquad \sigma = \frac{4\,y^3+9\,x^2}{y^{3/2}},
\]
where $v(\sigma)$ is a solution to $v_{\sigma\sigma} = -3^{-1}\,\sigma^{-1}\,v_\sigma$.
This equation is integrable in quadratures, its general solution reads
$v=c_1+c_2\,\sigma^{2/3}$, where $c_1$, $c_2=\mathrm{const}$. Therefore we have
\[
w= \frac{c_1\,y}{(4\,y^3+9\,x^2)^{5/6}}+\frac{c_2}{(4\,y^3+9\,x^2)^{1/6}}.
\]

\subsubsection{Solution (\ref{phi_4_plus_alpha_phi_1_invariant_solution}) with $\beta \neq 0$}
\hspace {1 pt}
\vskip 2 pt
\noindent
For solution (\ref{phi_4_plus_alpha_phi_1_invariant_solution}) in the case of $\beta \neq 0$ the
second equation of system (\ref{FHZ_system}) acquires the form
\[
w_{yy} = -(3\,\alpha\,\beta\,y\,\vert 2\,x-\beta\,y^2\vert^{1/2}-2\,\beta\,x+2\,\beta^2\,y^2-y)\,w_{xx}
\]
\[
\qquad \qquad
-(3\,\alpha\,\vert 2\,x-\beta\,y^2\vert^{1/2}+2\,\beta\,y-\beta^{-1})\,w_{xy}.
\]
After the change of
variables $x = \case{1}{2}\,(\tilde{x}^2+\beta\,\tilde{y}^2)$, $y=\tilde{y}$, $w = \tilde{w}$ and  dropping tildes we get
\[
w_{yy} = \beta\,w_{xx}+\frac{1-3\,\alpha\,\beta\,x}{\beta\,x}\,w_{xy}.
\]
This equation has a nontrivial symmetry   $w_y-\lambda\,w$, $\lambda =\mathrm{const}$. The corresponding re\-duc\-ti\-on $w=e^{\lambda\,y}\,v(x)$ yields ODE
$v_{xx} = \lambda\,(3\,\alpha\,\beta \,x -1)\,\beta^{-2}\,x^{-1}\,v_x+\lambda^2\,\beta^{-1}\,v$, which after the change of variables
$v=\frac{\lambda}{2\,\beta}\,(3\,\alpha\,x-\ln x)\,\tilde{v}$ and dropping tildes takes the form
\begin{equation}
v_{xx} = \frac{\lambda}{4\,\beta^2}\,\left(
\lambda\,(9\,\alpha^2+4\,\beta)
-\frac{6\,\alpha\,\lambda}{\beta\,x}
+\frac{\lambda-2\,\beta}{\beta^2\,x^2}
\right)\,v.
\label{ode_3}
\end{equation}
Analysis of this equation splits into two branches. The first one corresponds to the ca\-se  of $9\,\alpha^2+4\,\beta \neq 0$.
In this case the scaling
$\tilde{x}= \lambda \,\beta^{-1}\,(9\,\alpha^2+4\,\beta)^{1/2}\,x$ after dropping tildes gives Whittaker's equation
\begin{equation}
v_{xx} = \left(
\frac{1}{4}
-\frac{\kappa}{x}
+\frac{\lambda\,(\lambda-2\,\beta)}{4\,\beta^2\,x^2}
\right)\,v
\label{ode_4}
\end{equation}
with
$
\kappa = 3\,\alpha\,\lambda\,\beta^{-2}\,(9\,\alpha^2+4\,\beta)^{-1/2}$
and
$\mu = \pm \case{1}{2}\,(\lambda\,\beta^{-2}-1)$.   As it was shown in \cite{RamisMartinet}, this equation is integrable in quadratures
whenever $\pm \kappa \pm \mu -\case{1}{2} \in \mathbb{Z}$.   Therefore for each choice of $\alpha$ and $\beta$ there exists an infinite number of values for $\lambda$ such that equation (\ref{ode_4}) is integrable in quadratures.

The second branch corresponds to the case of $9\,\alpha^2+4\,\beta = 0$. Then equation (\ref{ode_3}) acquires the form
\[
v_{xx} = \left(
\frac{A}{x}
+\frac{B}{x^2}
\right)\,v
\]
with
$A=\case{27}{2}\,\alpha^3\,\tilde{\lambda}^2$, $B=\tilde{\lambda}\,(\tilde{\lambda}-1)$, and
$\tilde{\lambda} = \case{8}{81}\,\lambda\,\alpha^{-4}$.  After the change of variables
$v = \case{1}{2}\,A^{-1/4}\,\tilde{x}^{1/2}\,\tilde{v}$, $x = \case{1}{16}\,A^{-1}\,\tilde{x}^2$
and dropping tildes we have Bessel's equation
\begin{equation}
v_{xx} = \left(
\frac{1}{4}
+\frac{4\,B+\case{3}{4}}{x^2}
\right)\,v,
\label{ode_5}
\end{equation}
which is integrable in quadratures whenever $B = -\case{3}{16}+\case{1}{2}\,\left(n+\case{1}{2}\right)^2$,
$n\in \mathbb{Z}$, see \cite{Ritt1948}. So for each choice of $\alpha$ there exists an infinite number of values for $\lambda$ such that equation (\ref{ode_5}) is integrable in quadratures.

\subsubsection{Solution (\ref{phi_4_plus_alpha_phi_1_invariant_solution}) with $\beta = 0$}
\hspace {1 pt}
\vskip 2 pt
\noindent
The second equation of system (\ref{FHZ_system})
that corresponds to solution (\ref{phi_4_plus_alpha_phi_1_invariant_solution})
in the case of $\beta = 0$ has the form
\[
w_{yy} = (2\,\alpha\,x-2\,\alpha^2\,y^2+y)\,w_{xx}-(2\,\alpha\,y-\alpha^{-1})\,w_{xy}.
\]
After the change of
variables
$x=\frac{1}{8}\,\alpha^{-3}\,(2\,\tilde{x}+\tilde{y}^2+2\,\tilde{y}-1)$,
$y=-\frac{1}{2}\,\alpha^{-2}\,\tilde{y}$,
$w = \tilde{w}$
and  dropping tildes we get
\[
w_{yy} = 2\,(x+y)\,w_{xx}+w_{x}.
\]
This equation has a nontrivial symmetry   $w_x+w_y-\lambda\,w$, $\lambda =\mathrm{const}$.
The corresponding re\-duc\-ti\-on $w=e^{\lambda\,x}\,v(\tau)$ with $\tau = x+y$ yields ODE
$v_{\tau\tau} = -((4\,\lambda\,\tau+1)\,v_\tau+\lambda\,(2\,\lambda\,\tau+1)\,v)\,(2\,\tau-1)^{-1}$,
which after the change of variables
$v=\tilde{x}^{-\lambda/2-1/4}\e^{-\tilde{x}/2}\,\tilde{v}$,
$\tau = \frac{1}{2}\,(\lambda^{-1}\,\tilde{x}+1)$
and dropping tildes acquires the form of  Whittaker's equation
\begin{equation}
v_{xx} = \left(
\frac{1}{4}
+\frac{3\,\lambda+1}{4\,x}
+\frac{(2\,\lambda+1)^2}{8\,x^2}
\right)\,v
\label{ode_3_a}.
\end{equation}
From results of \cite{RamisMartinet} it follows that this equation is integrable in quadratures
whenever
$\lambda = 1 \pm 12\,n \pm (128\,n^2-32\,n+6)^{1/2}$, $n \in \mathbb{Z}$.
Therefore for each
choice of $\alpha$ there exists an infinite number of values for $\lambda$ such that
equation (\ref{ode_3_a}) is integrable in quadratures.

\subsubsection{Solution (\ref{phi_4_plus_alpha_phi_5_invariant_solution})}
\hspace {1 pt}
\vskip 2 pt
\noindent
In the case of solution (\ref{phi_4_plus_alpha_phi_5_invariant_solution}) we can put $\beta = 0$ without loss of generality.  Than the  second equation of system (\ref{FHZ_system}) takes the form
\[
w_{yy} = \alpha\,w_{xx}-\alpha^{-1}\,x\,w_{xy}.
\]
The nontrivial symmetry  $w_y-\lambda\,w$, $\lambda = \mathrm{const}$ leads to the reduction $w=e^{\lambda\,y}\,v(x)$, where $v$ is a solution of ODE
\[
v_{xx} = \lambda\,\alpha^2\,x\,v_x+\lambda^2\,\alpha\,v.
\]
After the change of variables
$v=e^{\frac{1}{4}\,\tilde{x}^2}\,\tilde{v}$, $x = \beta^{-1}\,\lambda^{-1/2}\,\tilde{x}$ and dropping tildes we have
Weber's equation
\begin{equation}
U_{xx} = \left(\case{1}{4}\,x^2+\mu-\case{1}{2}\right)\,U
\label{ode_6}
\end{equation}
with
$\mu = \beta^{-1}\,\lambda^{-1/2}$, which is  integrable in quadratures whenever $\mu \in \mathbb{Z}$, \cite{Kovacic1986}.
Therefore for each choice of $\alpha$ there exists an infinite number of values for $\lambda$ such that equation (\ref{ode_6}) is integrable in quadratures.

\section{Acknowledgments}
We are very grateful to Maxim V. Pavlov for important and stimulating discussions.

\end{document}